\documentstyle[11pt]{article}  
  
\begin{document}

\title{Group selection models in prebiotic evolution }

\author{D. Alves, P. R. A. Campos, A. T. C. Silva and 
J. F. Fontanari \footnote{corresponding author} \\
 Instituto de F\'{\i}sica de S\~ao Carlos \\
 Universidade de S\~ao Paulo \\
 Caixa Postal 369 \\
 13560-970 S\~ao Carlos SP, Brazil 
}

\date{}

\maketitle

\small 

\bigskip

\centerline{\large{\bf Abstract}} 

\bigskip

The evolution of enzyme production is studied analytically using
ideas of the group selection theory for the evolution of altruistic
behavior.
In particular, we argue that the mathematical formulation of 
Wilson's structured deme model ({\it The Evolution of Populations and
Communities}, Benjamin/Cumings, Menlo Park, 1980) is a mean-field
approach in which the actual environment that a particular individual 
experiences is replaced by an {\it average} environment. That
formalism is further developed so as to avoid the mean-field approximation
and then applied to the problem of enzyme production in the prebiotic context,
where the  enzyme producer molecules play the altruists role while
the molecules that benefit from the catalyst without paying its production
cost play the non-altruists role.  The effects of synergism  
(i.e., division of labor)  as well as of mutations are also considered
and the results of the equilibrium analysis are summarized in phase
diagrams showing the regions of the space of parameters where
the altruistic, non-altruistic and the coexistence regimes are stable.
In general, those regions are delimitated by discontinuous transition
lines which end at critical points.

\bigskip


\newpage


\section{Introduction}\label{sec:level1}


The controversial issue of the evolution and maintenance of altruism 
 has probably entered the field of prebiotic 
evolution when Maynard Smith \cite{Maynard} remarked that giving catalytic 
support in a 
molecular catalytic feedback network, such as the hypercycle \cite{hyper}, 
is in fact an altruistic behavior. As a result, 
such systems are extremely vulnerable to the presence
of parasites, i.e., molecules that  receive catalytic support
but do not give support to any other molecule in the network.
However, the stability of this type of cooperative networks
is crucial for the theories on the origin of life, as  
Eigen has shown  that the lengths of competing self-replicating molecules 
are limited by their replication accuracies and so they cannot
integrate sufficient information to code for a complex metabolism
\cite{Eigen,review}.
For the sake of concreteness, we define
an  altruistic behavior  as one that is detrimental to
the fitness of the individual who expresses it, but that confers an
advantage on the group of which that individual is a member \cite{altru}.
 
In the traditional group selection modeling, based on the 
Island models of Wright \cite{Wright}, it is assumed that the population is
divided into reproductively isolated subpopulations or
demes \cite{altru}.  The stability of the altruists
is achieved by postulating the existence of an external extinction mechanism 
acting on the demes that takes place at a rate depending on the deme
composition. Of course, such extinctions will  favor the occurrence of
individuals that lower the probability of extinction of the deme
they belong to which, in the case, are the altruistic individuals
 \cite{Eshel,Aoki,Ana}. A more modern formulation of group selection put
forward by Wilson \cite{Wilson} considers the demes  as
{\it trait groups}, in which the actual ecological, biochemical
or social interactions occur, but the  individuals 
are allowed to access and compete for the total resources available in 
the environment. Clearly, in this formulation the notion of group or deme
is somewhat blurred since, as will become clear in the examples given below, there
is a stage of the life cycle of the individuals when they leave their
demes to (effectively) interact with each other.

Actually, it is not so hard to envision physical systems
described by Wilson's trait group or structured deme model. 
For instance,  some basic
features of viral selection dynamics  can be modelled by viewing the 
cells as demes \cite{viral1,viral2}. In this case, it is assumed that
only $N$ free viruses 
entry and hence infect a cell; however inside the cell the viruses
undergo exponential growth leading to the burst of the cell and the
consequent release of free viruses which will again infect (colonize) 
other cells, and so on. As only $N$ viruses can infect each  cell,
there is an effective competition between all individuals in the
population. This restriction, though very far-fetched, does not
seem to change  qualitatively the behavior of the system
\cite{viral1} and, in addition, it suits very well to describing
{\it in vitro} serial
passage of viruses \cite{viral2}.
Other interesting application of Wilson's
formalism, which will be the main concern of this paper,
is the evolution of enzyme production in the prebiotic context
\cite{Michod,SM}. Here the demes are rock crevices or suspended 
water droplets of some fixed size. As before, although the 
macromolecules inside the demes undergo  exponential growth, they are
regularly washed away by tides or distributed by winds, and
only a small fraction of them is then re-adsorbed to the cracks or 
droplets. Both examples show that the spatial localization of 
viruses or macromolecules facilitates the selection against 
parasites. Henceforth we will refer to the individuals that
do not display altruistic behavior as non-altruists instead of
parasites, since 
they can subsist even in the complete absence of altruists.

The mathematical formulation of Wilson's structured deme model 
is centered on the concept of the average subjective frequencies
of altruists, which are defined as the frequencies of altruists experienced 
by the  {\it average} altruist and non-altruist in the population \cite{Wilson}. 
These quantities differ from the global frequency of altruists 
because the variance of the distribution of the deme 
compositions is non-zero, i.e., the population is not homogeneous. 
In particular, the stability of the altruists is achieved
by assuming that the fitness of both  altruists and non-altruists are 
proportional to their subjective frequencies. This formulation
may be viewed as a sort of mean-field approach in the sense that
the fitness of a given individual, say a non-altruist, in a particular 
deme is not proportional to the frequency of altruists it actually 
experiences (i.e., the fraction of altruists in its deme) but to 
the frequency of altruists
experienced by the average non-altruist in the population.
In this paper we show that going beyond this mean-field approach
does not make the theory any more complicated and, in addition,
it allows the identification of a recently proposed  model for the
evolution of altruism \cite{Roberta,DPS},  as well as of a population genetics
formulation of Eigen's model of molecular evolution \cite{Alves96},
as variants of Wilson's group selection model.

The remainder of the paper is organized as follows. In Sec.\
\ref{sec:level2} we present the general formalism that takes
into account that the fitness of an individual depends on the
fraction of altruists it actually experiences in its deme. Otherwise
the model conforms to Wilson's trait group model with the unlimited 
growth inside the demes followed by the  destruction of the demes,
and the random sampling of $N$ individuals to form each new deme. 
The formalism is then applied to the detailed study of a
model for the evolution of enzyme 
production proposed by Michod \cite{Michod} in Sec.\ \ref{sec:level3}. 
Building on the work of Donato \cite{Roberta,DPS}, in Sec.\ \ref{sec:level4}
we  apply our formalism to investigate the effects of 
synergism or division of labor in the prebiotic problem of enzyme 
evolution. A variant of the quasispecies model of molecular evolution
in which the replicating entities are the individual monomers that build up
the molecules is considered in Sec.\ \ref{sec:level5}.
Finally, some concluding remarks are presented in
Sec.\ \ref{sec:level6}.

%
\section{The model}
\label{sec:level2}
%

The population is composed of an infinite number of demes,
each of which is composed of $N$ haploid, asexually reproducing 
individuals. The individuals can be of two types, $A$ or $B$, 
depending on whether they present altruistic or non-altruistic
behavior, respectively. By definition, altruistic individuals
increase the fitness or reproductive rate of all individuals in 
the deme they belong to, but
pay a price for that by reducing their own fitness. 
Thus the key ingredient of any group selection model is that the fitness
of the individuals depend on the composition of the demes,
which are classified according to the number of altruists
they have: there are $N+1$ different types, labeled by the
integers $i=0,1,\ldots,N$. Hence, an  altruistic individual living
in a deme of type $i$ has fitness $F_A(i)$ while a  non-altruistic individual
living in the same deme has fitness $F_B(i)$, with $F_B (i) \geq F_A(i)$. 
Clearly, either in the viral dynamics or in the enzyme production problem
mentioned before, the  occurrence of errors in the replication of the
individuals (viruses or macromolecules) may have important
implications to the equilibrium composition  of the population. 
In order to take this possibility into account we  introduce the mutation 
rate $u \in [0,1/2]$, which gives the
probability that type  $A$ mutates to  type $B$ and vice-versa. 

To derive a recursion equation  for the frequency of altruists $p_t$ in
the population  at generation $t$ it is more convenient to
introduce the frequency of demes with  $i=0,\ldots,N$ 
altruists in generation $t$, denoted by
$Y_t \left( i \right)$. According to  the discussed above,
and assuming, as usual,  non-overlapping generations (i.e., all individuals
in generation $t$ are replaced  by their offspring in generation $t+1$)
the {\it average} number of altruists ${\mathcal N}_A$ and 
non-altruists ${\mathcal N}_B$ generated during
the stage of unlimited growth  inside the demes are
\begin{equation}
{\mathcal N}_A = \sum_{i=0}^N  
\left [ \left ( 1 -u \right) i F_A \left( i \right)  
+ u \left (N - i \right) F_B \left( i \right) \right ]  Y_t \left( i \right)
\end{equation}
and
\begin{equation}
{\mathcal N}_B = \sum_{i=0}^N  
\left [ \left ( 1 -u \right) \left (N - i \right)  F_B \left( i \right)  
+ u i F_A \left( i \right) \right ]  Y_t \left( i \right)
\end{equation}
respectively. Hence the global frequency of altruists in the
(free) population  at generation $t+1$ is given by
\begin{eqnarray}\label{pt}
p_{t+1} & = & \frac{{\mathcal N}_A}{{\mathcal N}_A+{\mathcal N}_B}
\nonumber \\
& = & u + \frac{1-2u}{w_t} \sum_i i F_A \left( i \right) Y_t \left( i \right)
\end{eqnarray}
where
\begin{equation}\label{wt}
w_t = \sum_{i=0}^N  \left [ i F_A \left( i \right) + \left( N - i \right)
F_B \left( i \right) \right ] 
Y_t \left( i \right)
\end{equation}
is the average fitness of the population.
The next step in modelling  is to
distribute these individuals in infinite demes, each of which containing
exactly $N$ individuals. In the absence of additional information, the most
conservative assumption that can be made about the re-grouping mechanism 
is that the individuals are picked randomly from the (free) population.
This leads to the binomial distribution  
\begin{equation}\label{bin}
Y_{t+1} \left( i \right) = \left ( \! \! \begin{array}{c} N \\ i \end{array}
      \! \! \right ) \,
\left ( p_{t+1} \right )^{i} \, \left  (1 - p_{t+1} \right )^{N -i},
\end{equation}
which together with Eqs.\ (\ref{pt}) and (\ref{wt}) allow the complete
description of the life cycle of the individuals. 

For the sake of completeness and to facilitate comparisons
between the two formalisms, at this point  it is
convenient that we introduce the basic ingredients of 
the original structured deme formalism as proposed by Wilson \cite{Wilson}.
The conditional probability distributions
of type $A$ given type $l=A,B$ at generation $t$ are defined by
\begin{eqnarray}
{\mathcal{P}}_t \left ( i | A \right) & = &  \frac{i Y_t \left( i \right)}
{\sum_{i=0}^N i Y_t \left( i \right)} \label{P_A} \\
{\mathcal{P}}_t \left ( i | B \right) & = &  \frac{\left (N - i \right)  Y_t \left( i \right)}
{\sum_{i=0}^N \left (N - i \right) Y_t \left( i \right)} \label{P_B} 
\end{eqnarray}
which must be interpreted as follows: considering a particular replicator of type $l$
then ${\cal{P}}_t \left ( i | l \right)$ is  the probability that such a replicator
belongs to a deme containing $i$ individuals of type $A$.
Hence the average
subjective frequency of altruists as seen by altruists is given by
\begin{equation}\label{f_A}
f_A (t) = \frac{1}{N} \sum_{i=0}^N i {\mathcal{P}}_t \left ( i | A \right) = 
p_t + \frac{\sigma_t^2}{N^2 p_t}
\end{equation}
where $\sigma^2_t = \sum_i i^2 Y_t (i) - \left[ \sum_i i Y_t (i)\right]^2$ is the
variance of the deme distribution and $p_t = \sum_i i Y_t (i) /N$ is the global
frequency of altruists in the population. 
Similarly, the average subjective frequency of altruists as seen by non-altruists
is
\begin{equation}\label{f_B}
f_B (t) = \frac{1}{N} \sum_{i=0}^{N} i {\mathcal{P}}_t \left ( i | B \right)
= p_t - \frac{\sigma_t^2}{N^2 \left ( 1 - p_t \right )} .
\end{equation}
In the case the demes are assembled randomly, obeying a binomial distribution,
one has $\sigma_t^2 = N p_t \left (1 - p_t \right)$ so that for large $N$
the subjective frequencies tend to the global one. 
Since
$f_A (t) \geq p_t \geq f_B (t)$, the main point of introducing the subjective 
frequencies is to show that a population structured in groups 
of distinct compositions can simultaneously 
enhance 
the  effects of the presence of altruists on themselves and  
diminish those beneficial effects on the non-altruists. 
Of course, the assumption that the distribution of deme compositions $Y_t (i)$
affects the dynamics only through the average subjective frequencies $f_l (t),
l=A,B$ is too restrictive, limiting, for instance, the choices for the 
dependence of the fitness  of the individuals on the deme composition.

%
\section{Evolution of enzyme production}\label{sec:level3}
%

According to the scenario proposed by Michod \cite{Michod}, we consider
two  types of replicators, $A$ and $B$, and assume that only replicator
$A$ can produce a catalyst (enzyme) which, however, can catalyze 
the replication of both types of replicators, but with different efficiencies.
Since 
replicator $A$, which produces the catalyst, must suffer some cost in its 
noncatalyzed self-replication rate, while replicator $B$ attains all
the benefits of the catalyst without paying the cost for its production,
we have here a typical situation of altruistic behavior.   
The cost associated with being altruistic is modelled by assigning the 
noncatalyzed self-replication  rate $1 - r$, with $r \in [0,1]$, to $A$ and 
the rate $1$ to $B$. Moreover, the rate of catalyzed replication is
proportional to the concentration of enzymes in the deme,
which in turn is proportional to the
concentration of replicators $A$ in that deme. Hence,
assuming that self-replication and the
replication catalyzed by the enzyme are separate processes, the
fitness of a replicator $l=A,B$ belonging to a deme of type $i$ 
can be written as
\begin{equation}
F_l \left ( i \right ) = 1 - \alpha_l r + k_l \frac{i}{N} 
~~~i = \alpha_l, \alpha_l + 1, \ldots,N-1+\alpha_l ,
\end{equation}
where $\alpha_l = 1$ if $l=A$ and $0$ if $l=B$. Here the parameters
$k_l$ represent the beneficial effect of enzyme mediated replication.
In particular, $k_B=0$ implies that the enzyme is specific for the
replicator which produced it, as in the one-membered hypercycle \cite{hyper}.
However, it seems more plausible to assume that the primordial enzymes
were some kind of general catalysts which would facilitate the replication
of a wide spectrum of replicators, so in the following we will assume
that $k_A \geq k_B$. We note that Michod considers the case $k_A=k_B$
and $u=0$ only \cite{Michod}.
The recursion equation (\ref{pt}) thus becomes
\begin{equation}\label{rec_mich}
p_{t+1} = u + \left ( 1 - 2 u \right )\frac{ 
p_t \left ( 1 - r \right )+ k_A p_t^2 + \frac{1}{N} k_A p_t \left ( 1 - p_t \right ) }
{ 1 + p_t \left ( k_B - r \right )  + \left ( k_A - k_B \right ) 
\left [ \frac{1}{N} p_t \left ( 1 - p_t \right ) + p_t^2 \right ]},
\end{equation}
which is identical to that obtained  using 
Wilson's original (mean-field) formulation  \cite{Michod}. 
In fact, we note that the coefficient of $k_A$
in the numerator of Eq.\ (\ref{rec_mich}) can be written as $p_t f_A(t)$, 
so that the
rate of increase of altruists in the population due to the replication catalyzed
by the enzymes is proportional to the  average subjective
frequencies of altruists.

\begin{figure}
\vspace{9truecm}
\includegraphics{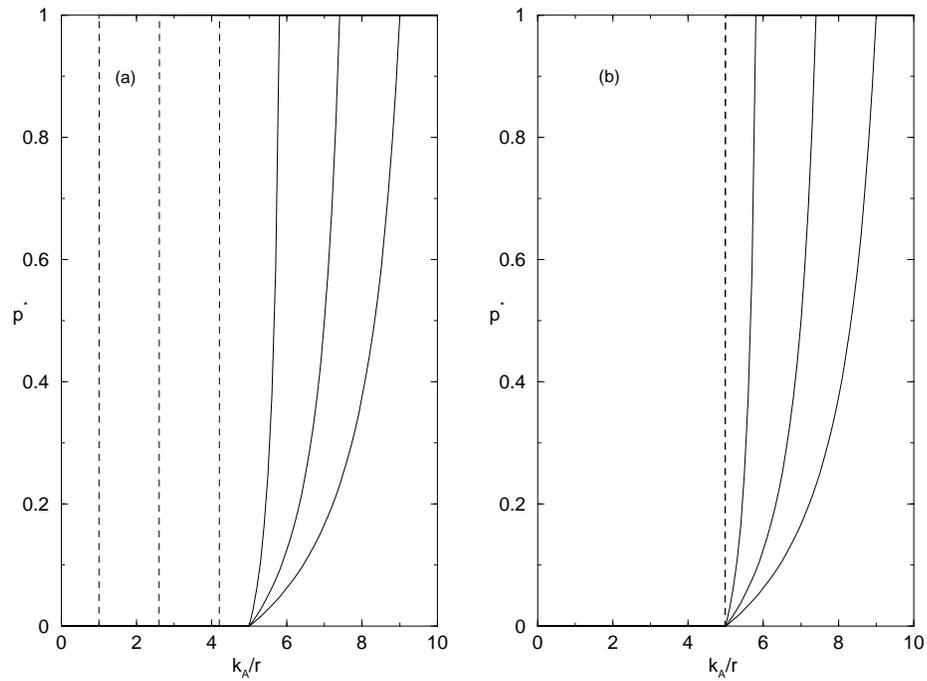}
\caption{Steady-state frequency of type $A$ replicators  for
$u=0$, $N=5$, and (from left to right) $k_{B}/r=0$, $2$, $4$, $6$, $8$ and $10$.
The initial frequencies are (a) $p_0 = .999$ and (b) $p_0 = 0.001$. The first three
lines in part (a) collapse into a single line in part (b). }
\label{fig:Mich1}
\end{figure}

\begin{figure}
\vspace{9truecm}
\includegraphics{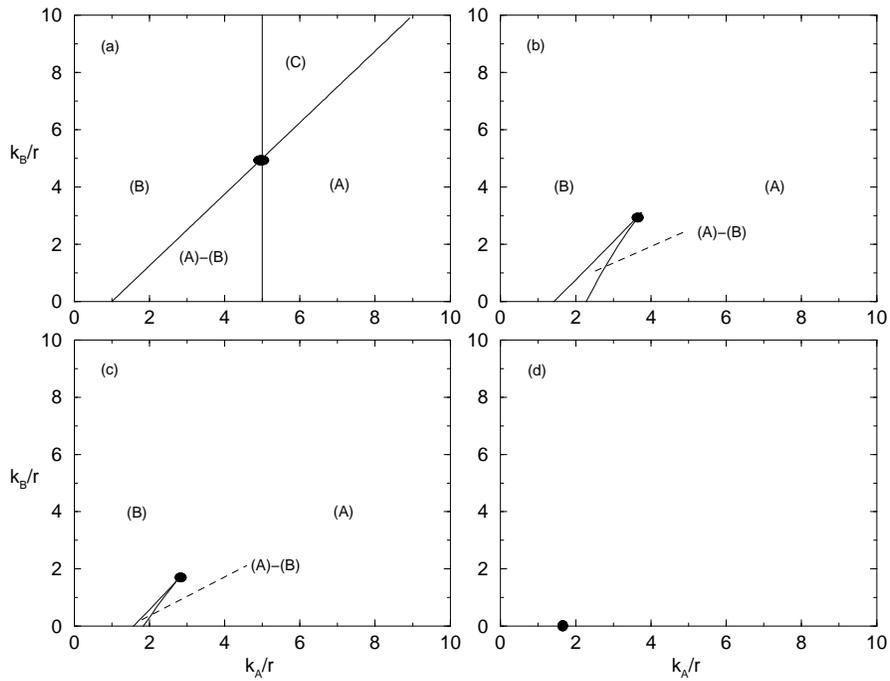}
\caption{Phase diagrams for $N=5$ and $r=0.1$ showing the regions of stability
of the different fixed points for (a) $u$=0, 
(b) $u$=0.005, (c) $u$=0.01 and (d) $u$=0.0158. The intersection point
touches the coordinate axis at $k_A/r = 5/3$.}
\label{fig:Mich2}
\end{figure}

It is instructive to consider first the case where mutations are not
allowed ($u=0$) 
since the steady-state equation obtained by setting $p_{t+1}=p_t = p^*$ 
can be solved analytically in this case. Explicitly, we find three 
fixed points: $p^*=0$, $p^*=1$ and
\begin{equation}\label{p_int}
 p^* = \frac{ r - k_A/N }{ \left ( k_A - k_B \right)
\left ( 1 - 1/N \right )} .
\end{equation}
A physically meaningful fixed point must be in the simplex $[0,1]$ and
satisfy the standard stability condition
\begin{equation}\label{stab}
\left. \frac{d p_{t+1}}{d p_t} \right |_{p_t = p^*}  < 1 .
\end{equation}
We find that $p^* = 0$ is stable for $k_A/r < N$, while $p^* = 1$
is stable for $k_A/r > 1 + \left ( 1 - 1/N \right ) k_B/r$. Interestingly,
for $k_B/r > N$ there is a region where both fixed points are unstable and
so the stable one is the  intermediate fixed point (\ref{p_int}) which
corresponds to a regime of coexistence between altruists and
non-altruists. These distinct regimes are illustrated
in Fig.\ \ref{fig:Mich1} where we show the steady-state frequencies
$p^*$ for two different  values of the initial frequency of altruists.  
We note that
in the case $u=0$ the analysis is considerably simplified as
only the ratios $k_l/r, ~l=A,B$ matter for the stability of the
fixed points. We identify four different phases
in the steady-state regime: the pure altruistic phase $(A)$ associated 
to the fixed point $p^*=1$; the pure non-altruistic phase $(B)$ 
associated to the fixed point
$p^*=0$; the coexistence phase $(C)$ associated to the fixed point
(\ref{p_int}); and the phase labeled $(A)\!-\!(B)$ where both
$p^*=1$ and $p^*=0$ are stable. In this phase the two kinds of replicators
compete such that there is an all-or-none selection, though the winner 
is not determined by the fitness only, but also by its initial abundance in
the population. In fact, the basins of
attraction of the two stable fixed points
are delimited by the intermediate fixed point (\ref{p_int}). These
results  are conveniently summarized in a 
phase diagram in the plane $\left ( k_A/r, k_B/r \right )$ as shown 
in Fig.\ \ref{fig:Mich2}(a). We note that the transitions between phases $(B)$
and $(C)$ as well as between phases $(C)$ and $(A)$ are continuous, in the sense
that $p^*$ increases continuously as those transition lines are crossed. 
It is important to note that
even in the case of completely nonspecific catalysis $k_A = k_B$
the altruistic replicators can dominate the entire population provided
that the condition $ k_A > r N$ is satisfied.

\begin{figure}
\vspace{9truecm}
\includegraphics{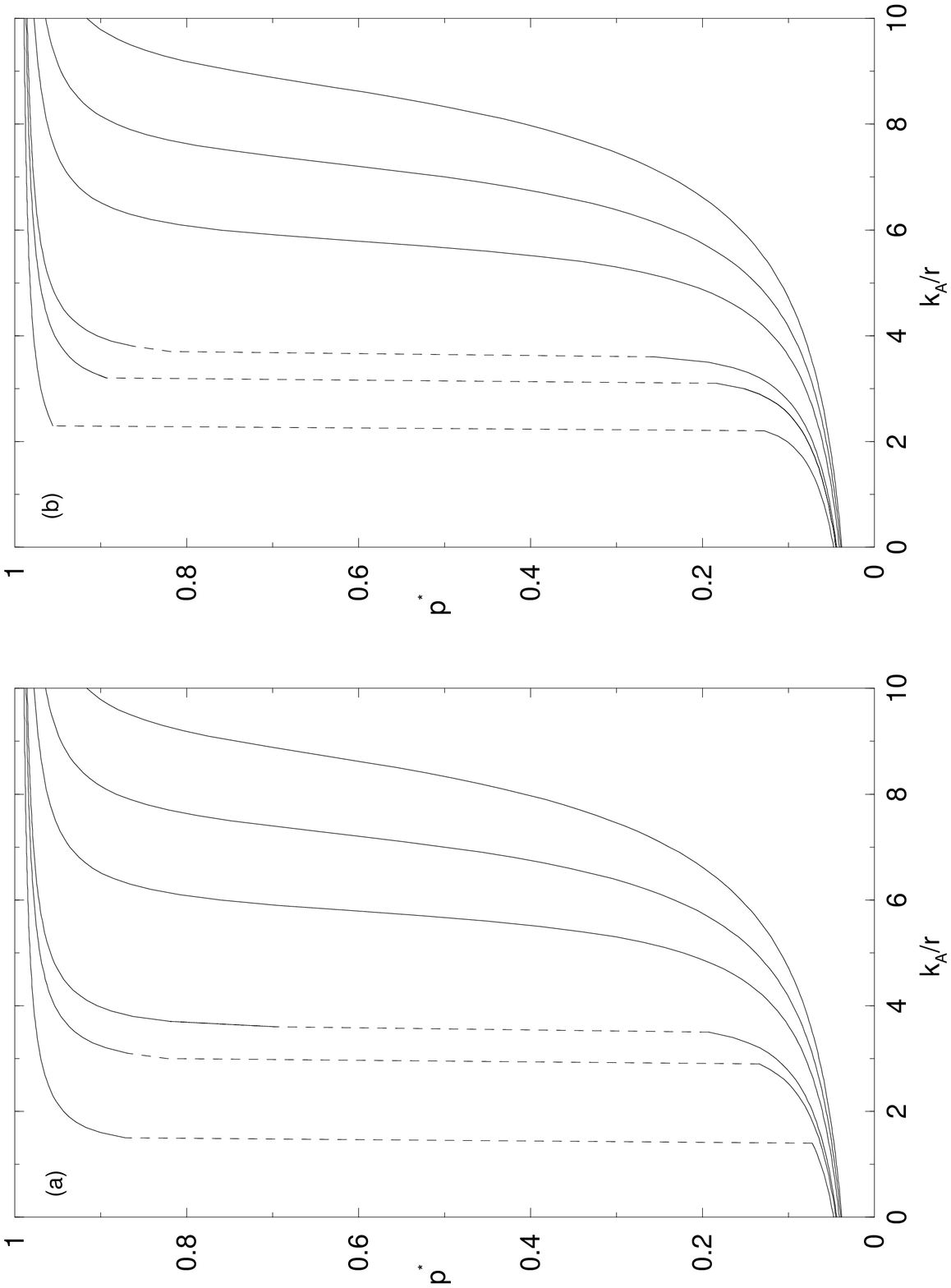}
\caption{Steady-state frequency of type $A$ replicators in the population for
$u=0.005$, $N=5$, $r=0.1$, and (from left to right) $k_{B}/r=0$, $2$, $2.9$, $6$,
$8$ and $10$. The initial frequencies are (a) $p_0 = 1$ and (b) $p_0 = 0$.}
\label{fig:Mich3}
\end{figure}

\begin{figure}
\vspace{9truecm}
\includegraphics{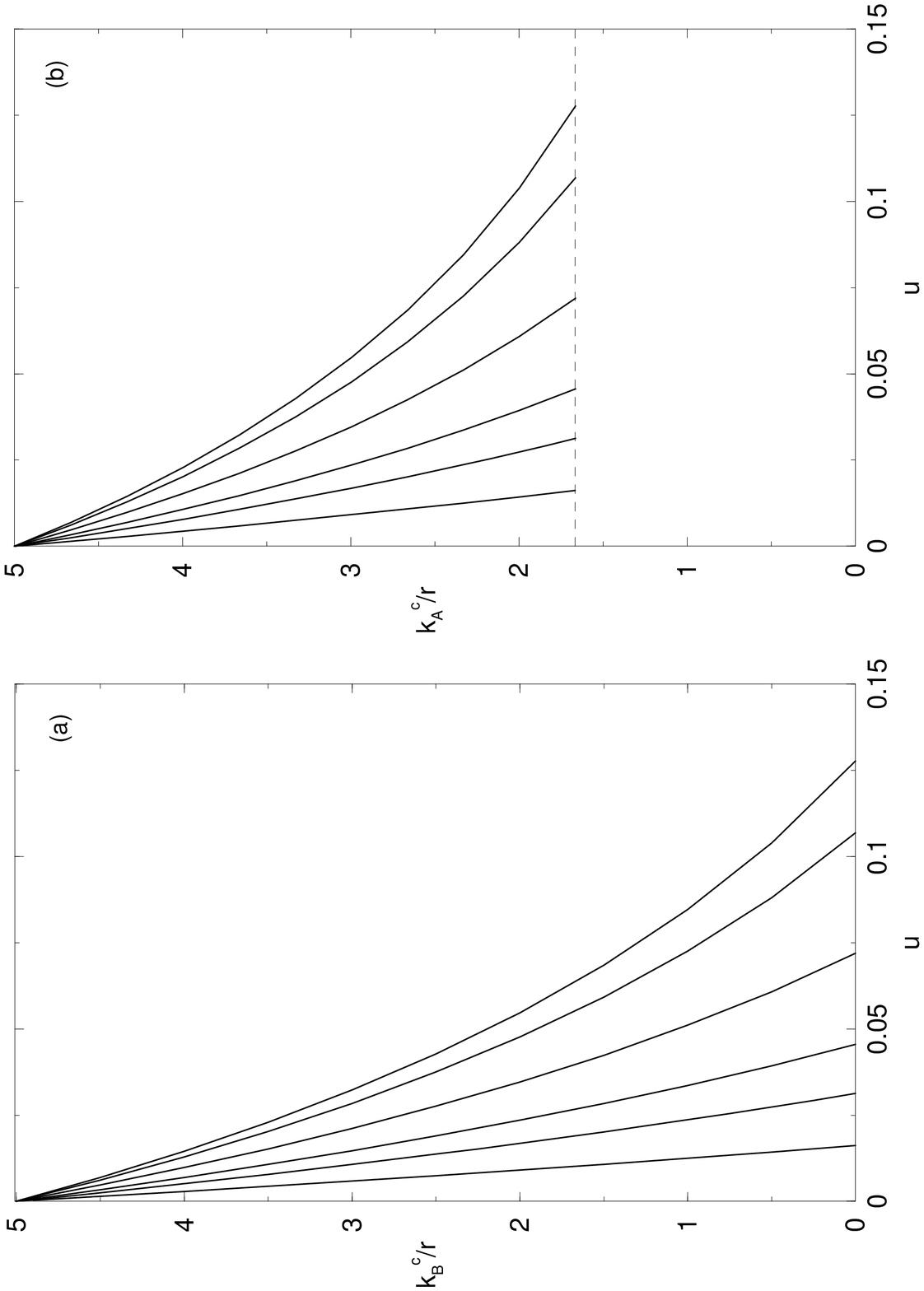}
\caption{Coordinates of the critical point  (a) $k_B^c/r$ and (b) $k_A^c/r$ 
as functions of the mutation rate $u$ for $N=5$ and (from left to right) 
$r=0.1$, $0.2$, $0.3$, $0.5$, $0.8$, and $1.0$. At the points where
$k_B^c/r =0$ we find $k_A^c/r = 5/3$.}
\label{fig:Mich4}
\end{figure}

\begin{figure}
\vspace{9truecm}
\includegraphics{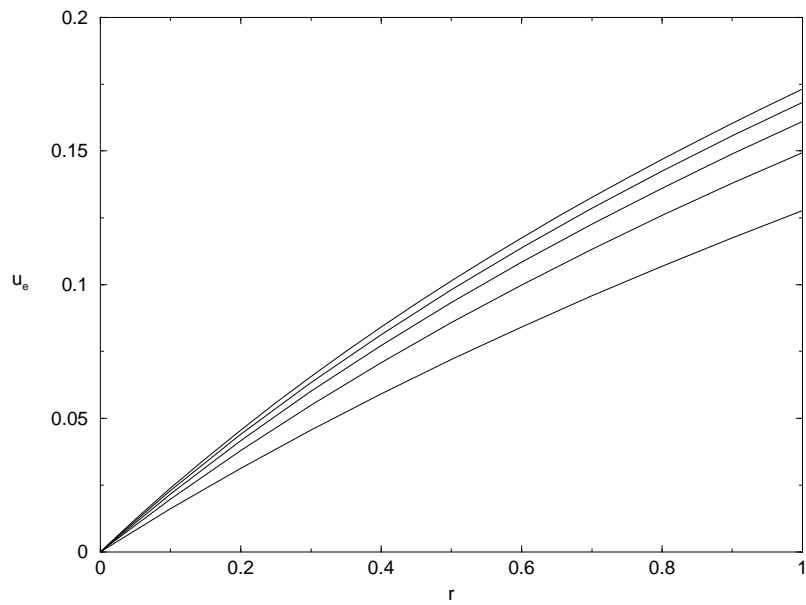}
\caption{Mutation rate $u_e$ beyond which the discontinuous transitions disappear
as function of the altruistic cost $r$ for 
(from bottom to top) $N=5$, $10$, $20$, $50$ and $\infty$.}
\label{fig:Mich5}
\end{figure}

We turn now to the more general case where the mutation rate $u$ is
nonzero. The obvious complication in this case is that $p=0$ and
$p=1$ are no longer fixed points and so, in principle, the phases
identified before cannot be unambiguously defined. 
However, the threshold phenomena observed in the dependence of the
steady-state frequency of type $A$ replicators on the 
scaled catalyst specificity $k_A/r$ (see Fig.\ \ref{fig:Mich3}) 
indicates that an unique extension of the definitions of
phases  $(A)$, $(B)$ and $(A)\!-\!(B)$ is possible indeed,
provided that $k_B/r$ is not larger than some critical value.
As expected, phase $(C)$ disappears since its defining characteristic,
namely, $0 < p^* < 1$, occurs for all parameter settings in the case 
of nonzero mutation rates.
The rich interplay between 
the stable fixed points is illustrated in the phase diagrams of
Fig.\ \ref{fig:Mich2}. The prominent feature of those phase diagrams 
is the existence of  {\it critical points} at which the two discontinuous
transition lines intersect and, as a result, above which it is no longer possible
to distinguish between phases $(A)$ and $(B)$. For fixed $u$, $r$ and $N$
the critical point coordinates $(k_A^c,k_B^c)$ are determined by requiring that
the three fixed points of the recursion equation (\ref{rec_mich})
collapse into a single one. Accordingly,  in Fig.\ \ref{fig:Mich4}
we show  the critical point coordinates as function of the mutation rate $u$. 
As expected, for $u=0$ 
we find $k_B^c/r = N$ regardless of the value of $r$. Of particular interest
is the mutation rate at which $k_B^c$ vanishes, henceforth denoted by $u_e$, as it
signalizes the disappearance of all traces of the two distinct regimes
associated to altruistic and non-altruistic behaviors,
leading to the phase diagram
of Fig.\ \ref{fig:Mich2}(d). Interestingly, at this value of the mutation
rate we find   $k_A^c/r = 2N/\left (N+1 \right)$ independently of $r$.
The dependence of $u_e$ on the altruistic cost $r$ for several values of 
the deme sizes is illustrated in Fig.\ \ref{fig:Mich5}.

The importance of the finitude of the
deme sizes $N$ to the stabilization of the altruists can be
appreciated by considering the limit $N \rightarrow \infty$,
which corresponds to a homogeneous population,
in the case of absence of mutations $u=0$.
In fact, in this case
the fixed point $p^*=0$ is always stable, while 
$p^*=1$ becomes stable only for $k_A > r + k_B$, which is a very
uninteresting situation from the point view of the evolution of altruism
since  in this case the effective fitness of an altruist 
($1 - r + k_A$) 
is larger than the fitness of a non-altruist ($1 + k_B$) belonging to the same deme.

%
\section{Synergism}\label{sec:level4}
%

A puzzling problem in evolution is the existence of complex structures
that are of value to the organism only when fully formed  \cite{Maynard2}.
It might be possible that enzyme
production has become a reality due to the combined work of  several
molecules, each being responsible for the synthesis of different pieces
of the catalyst. This situation of division of labor between
the altruists, termed synergism,  can result in highly non-additive fitness
interactions. To model this case, we assume that  the advantage to the deme 
is accrued only if the  number of altruists reaches  some minimal value. 
Explicitly, we will assume that only
individuals belonging to demes composed of $i \geq i_m$, with
$i_m= 0,1, \ldots,N$,   altruists 
have their fitness enhanced: for such demes
all individuals have their fitness increased by the factor
$1/\left ( 1 - c \right )$ with $c \in [0,1]$. 
The dependence of the  fitness  of types $A$ and $B$ on the composition 
of the deme is summarized by the following equation  
\begin{equation}\label{fit_syn}
F_l \left (i \right ) = \left \{ \begin{array}{ll}
				  1 - \alpha_l r & \mbox{ if $i<i_m$} \\				  
         \left ( 1 - \alpha_l r \right )/\left ( 1 - c \right ) & \mbox{otherwise}
				\end{array}
			\right.
\end{equation} 
where, as before, $r \in (0,1)$ is the cost for being altruistic and
$\alpha_l = 1$ if $l=A$ and $0$ if $l=B$.
The recursion equation (\ref{pt}) is then written as
\begin{equation}\label{rec_syn}
p_{t+1} = u + \left ( 1 - 2 u \right) 
\frac{ \left ( 1 - r \right ) \left [ p_t \left ( 1 - c \right )
+ \frac{1}{N} \, c \sum_{i=i^*}^N i Y_t \left ( i \right ) \right ]}
{\left ( 1 - c \right )\left ( 1 - r p_t \right ) + 
c \sum_{i=i^*}^N  Y_t \left ( i \right ) \left ( 1 - \frac{1}{N}  r i \right )} .
\end{equation}
The formalism based on the average subjective frequencies
cannot be applied to describe this dynamics because of the highly nonlinear
dependence of the fitness $F_l$  on the number of altruists in the deme.

Before we proceed with the analysis of the steady-state solutions of 
recursion equation (\ref{rec_syn}), we must  note that the fitness assignment 
summarized in
eq.\ (\ref{fit_syn}) was used  by Donato \cite{Roberta} in an 
alternative model for the selection of altruistic behavior,
which, similarly to Wilson's structured deme model, though not explicitly
acknowledged by that author, has a stage of
the life cycle of the individuals when they interact with all 
other individuals in the population. In fact, this must be so  because 
in Donato's model the relative 
fitness of an individual, which is related to
the number of offspring it generates, is defined as the ratio 
between the fitness of that individual and the fitness of the
{\it whole} population \cite {Roberta,DPS}. However, that model
has two other ingredients that differ  from Wilson's:
(i) the sizes of the demes are not fixed {\it a priori}, but there is a maximal deme 
size that once reached leads the deme to split in two smaller ones; and
(ii) the offspring of the individuals of one deme in one generation form one
deme in the next generation. These rules were motivated by the analogy
with social animals which live in groups not too large and whose offspring
remain in the group of their parents.
An interesting outcome of the model is the possibility of stable coexistence between
altruists and non-altruists within a same group, which is in fact the
situation observed in nature since  the altruistic behavior is usually  exhibited 
only by some individuals in the group. This result contrasts with that predicted
by the Island group selection models, namely, that in the absence of mutations
there are either fully altruistic  ($i=N$) or fully non-altruistic ($i=0$) groups
only \cite{Eshel,Aoki,Ana}. As we will show in the sequel, using the fitness assignment
of eq.\ (\ref{fit_syn})  this coexistence regime is associated to one of 
the stable steady-state solutions of the recursion equation (\ref{rec_syn}).

As before, we will consider first the simpler case where $u=0$. 
As expected,  $p=1$ and $p=0$ are always fixed points and, depending
on the values of the control parameters $i_m$, $c$ and $r$, there
can be either one or two additional fixed points. 
On the one hand, $p^*=0$ is
always stable for $i_m > 1$, while for $i_m =1$ it becomes  unstable 
in the region $c > r$. In fact,  for fixed $r$ a stable fixed point 
appears at $c=r$,
increasing continuously from $0$ to $1$ as $c$ increases. This 
behavior signalizes  the occurrence
of a continuous transition  from a regime characterized by fully non-altruistic
demes only ($p^*=0$) to a regime where inhomogeneous demes 
 formed of
both altruistic and non-altruistic individuals are allowed also ($ 0 < p^* < 1$). 
On the other hand, $p^*=1$ is always unstable for $i_m < N$, while for $i_m=N$
it becomes stable in the region $c>r$. In this case both fixed points
$p^*=1$ and $p^*=0$ are stable but, as pointed out before, only one of the
two types of individual will take over the population. 
In the other cases $1 < i_m < N$ the intermediate, stable fixed point
$ 0 < p^* < 1$ appears in a discontinuous manner, i.e., $p^*$ is non-zero
already at the outset.

\begin{figure}
\vspace{9truecm}
\includegraphics{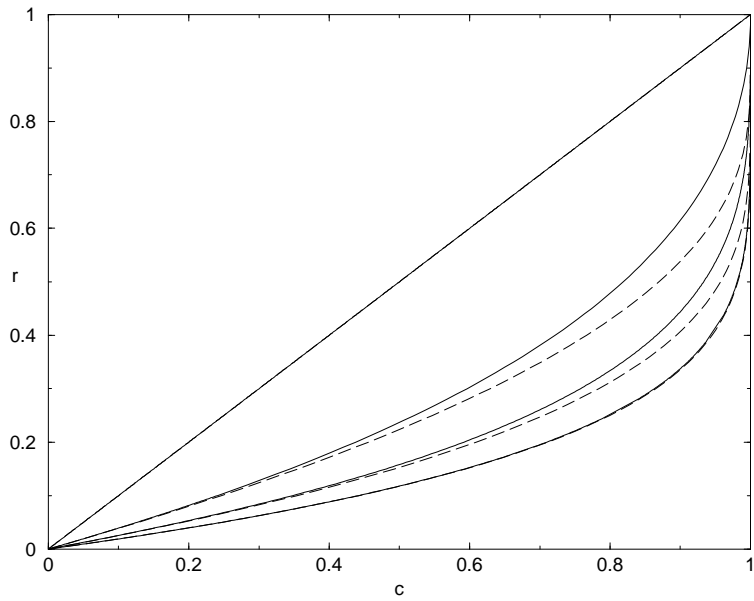}
\caption{Transition lines for $N=20$,  $u=0$ and, from top to bottom,
$i_m=1$, $2$, $4$, $10$ (solid lines), and $i_m=19$, $17$ (dashed lines).  
The curves for $i_m=20$ and $i_m= 11$ coincide with those for 
$i_m=1$ and $i_m= 10$, respectively.}
\label{fig:syner0}
\end{figure}

In Fig.\ \ref{fig:syner0} we present the transition lines separating
the region in the plane $(c,r)$ where the altruistic individuals persist
in the population (region below the curves) from the region where the
only stable fixed point is  the non-altruistic one $p^*=0$.
Since those curves satisfy $c \geq r$, it seems that the surviving altruists
are those belonging to demes with $ i \geq i_m$ since they have a larger fitness
than non-altruists living in  demes with $i < i_m$. Interestingly, 
the size of the region of existence of altruists  decreases with increasing 
$i_m$, reaches a minimal value for  $i_m = N/2$, and then  increases towards its
initial size as $i_m$ approaches $N$ (the transition lines for $i_m=1 $ and 
$i_m=N $ coincide).
However, it must be noted that the most favorable situation  to the altruists
is the case of no synergism $i_m = 1$, since only then
the fixed point $p^*=0$, associated to the non-altruistic regime, 
 becomes unstable. Moreover, the basin of attraction of the
intermediate fixed point decreases with increasing $i_m$ and so, 
unless there is already a large number of altruists at the outset, the
non-altruists will take over the population.
For instance, for $i_m =N$ the basin of attraction of $p^*=1$ is vanishingly
small close to the transition line $c=r$. This rather frustrating result
simply reflects the difficulty, already 
pointed out in the beginning of the section,
of evolving a synergistic system in nature. A possible solution to this
problem is provided by the so-called Baldwin effect \cite{Maynard2}
which, in the framework proposed by Hinton \& Nowlan \cite{HN,FM},
assumes the existence of a third
type of individual, say $X$,  which  either by learning, guessing
or imitation
can act as an individual of type $A$ or $B$ but whose offspring are,
of course, of type $X$. These plastic  individuals may  provide
the appropriate conditions (i.e., a large  number
of altruistically behaving individuals)  to start the synergistic effects
and, once this is done, they will become extinct due to the competition
with the born altruists, leaving thus no trace of
their early presence in the population. We will leave the investigation of
this avenue of research for a future contribution.

\begin{figure}
\vspace{9truecm}
\includegraphics{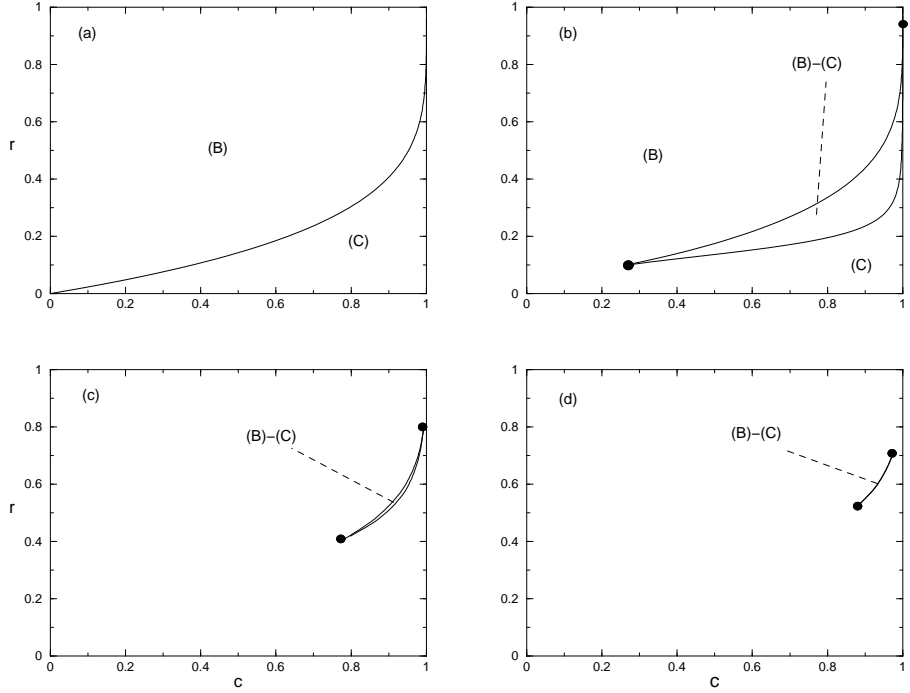}
\caption{Phase-diagrams for $N=20$ and $i_m=5$ showing the regions of stability
of the different fixed points for (a) $u =0$ , (b) $u =0.0085$ , (c) 
$u = 0.0282 $ and (d) $u = 0.0330$.}
\label{fig:syner1}
\end{figure}

\begin{figure}
\vspace{9truecm}
\includegraphics{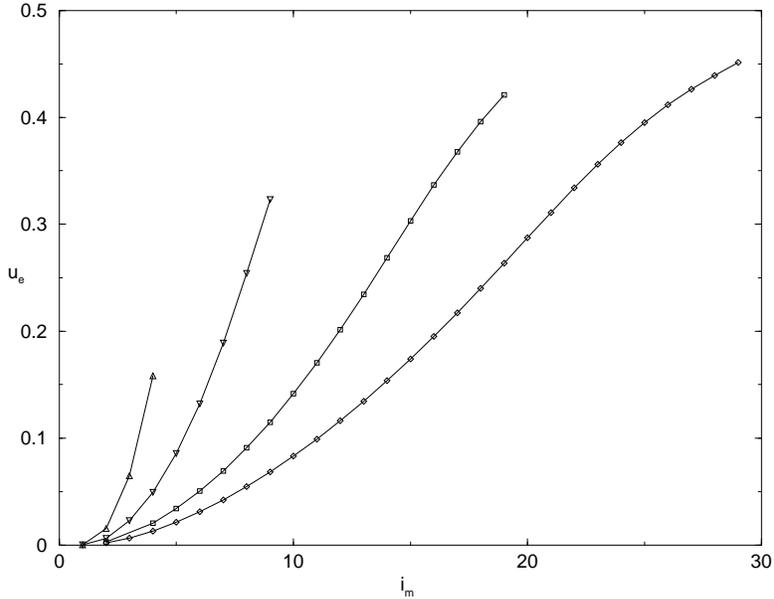}
\caption{Mutation rate $u_e$ beyond which the discontinuous transitions disappear
as function of $i_m$ for  $N=5$ ($\bigtriangleup$),
$N=10$ ($\bigtriangledown$),
$N=20$ ($\Box$), and 
$N=30$($\Diamond$). The lines are guides to the eye.}
\label{fig:syner2}
\end{figure}

Taking into account the effect of mutation ($u> 0$) leads to a very rich
interplay between the different steady-state regimes  of the recursion
equation (\ref{rec_syn}) as 
illustrated by the phase-diagrams shown in Fig.\ \ref{fig:syner1}.
In the absence of mutation
the phase labeled $(B)$ is associated to the non-altruistic
regime characterized by the fixed point $p^* = 0$, while  phase
$(C)$ is associated to the coexistence regime characterized by the intermediate
fixed point  $0 < p^* < 1$. As before, although for non-zero mutation rates
$p=0$ is no longer a fixed point it is still possible to distinguish between
the fixed points corresponding to the non-altruistic and the coexistence
regimes, due to the occurrence of threshold phenomena similar to those shown in 
Fig.\ \ref{fig:Mich3}. The main effect of mutation is to produce,  at the expense
of phase $(C)$,
a  bounded region, labeled $(B)\!-\!(C)$, where both phases are stable. This 
region is delimited by two discontinuous transition lines that intersect and end
at two critical points. As the mutation rate $u$ increases the size of the
bounded region decreases and disappears altogether at the critical end
point $u_e$ at which the two critical points coalesce. Hence 
for $u \geq u_e$ it is no longer possible
to distinguish between phases $(B)$ and $(C)$.
The dependence of $u_e$ on
$i_m$  is shown in Fig.\ \ref{fig:syner2}. As expected $u_e =0$ for $i_m=1$,
regardless of the value of the deme size $N$, since the transition between those
two phases is  continuous already for $u=0$.

%
\section{Quasispecies model}\label{sec:level5}
%

\begin{figure}
\vspace{9truecm}
\includegraphics{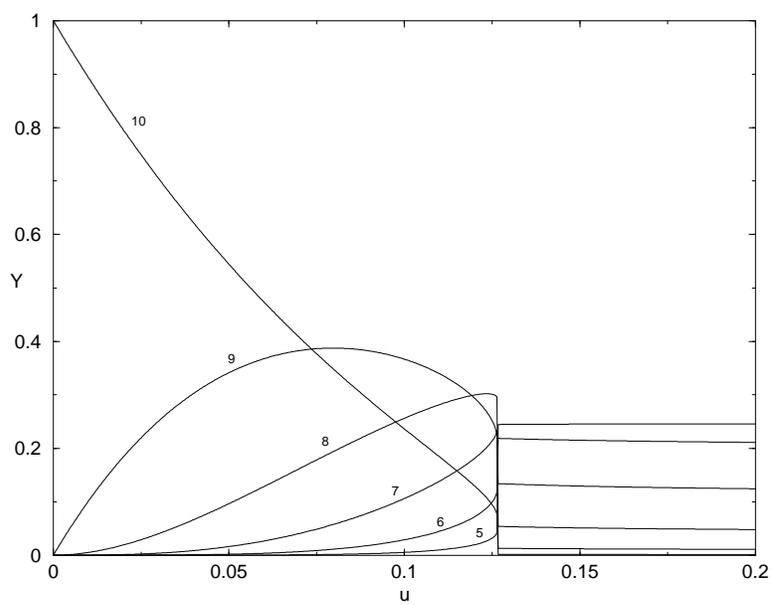}
\caption{Steady-state frequencies of molecules composed of $10$ (master), $9$,
$8$, $7$, $6$ and $5$ monomers of type $A$ as functions of the
error rate for $N=10$ and $s=0.9$. The initial type $A$ monomer frequency is
$p_0=1$.}
\label{fig:quasi1}
\end{figure}

Other interesting application of the formalism presented in 
Sec.\ \ref{sec:level2} is the study of the error threshold transition
in Eigen's molecular quasispecies model \cite{Eigen}. 
Such  approach has recently been proposed as an (uncontrolled) approximation 
to the original kinetics formulation  of the quasispecies model \cite{Alves96},
without the realization of its close connection with Wilson's trait group
framework. 
In this case,
the monomers play the role of the individuals, and the molecules  the
role of the demes. However, there is no distinction between altruistic
and non-altruistic monomers (i.e., there is no altruistic cost)
but the self-replication rates of the monomers depend on 
the molecule they belong to. Thus, contrasting to
Eigen's original proposal,  in this formulation
the molecules are not self-replicating entities,  being 
only passive carriers of  monomers. In this context, it is more appropriate
to think of the mutation rate $u$ as the replication error rate per monomer. 
Explicitly, for  the single-sharp-peak replication  
scenario we have $F_B \left ( i \right ) = 1-s, i = 0, \ldots, N-1$ and
\begin{equation}\label{fit_eig}
F_A \left (i \right ) = \left \{ \begin{array}{ll}
				  1-s  & \mbox{if $i=1,\ldots,N-1$} \\				  
         			1 & \mbox{if $i=N$}
				\end{array}
			\right.
\end{equation} 
where $0 \leq s \leq 1$ is the selective advantage of the so-called 
master molecule, namely, the molecule composed of $N$ monomers of type $A$
\cite{review}. 
The general recursion equation (\ref{pt}) then becomes
\begin{equation}\label{rec_quasi}
p_{t+1} = u + \left ( 1 - 2u \right ) \frac{\left (1 -s \right ) p_t +  s p_t^N}
		{1 - s + s p_t^N} .
\end{equation}
The only stable fixed point for $u=0$ is $p^* = 1$ which corresponds to
the domination of the population  by the master molecules. As the mutation rate
increases, two distinct regimes are observed in the composition of the
population: the {\it quasispecies} phase $(Q)$ characterized by the master molecule
and its close (in the sense of the Hamming distance) neighbors, and the
{\it uniform}  phase $(U)$ where the $2^N$ molecules appear in the same proportion.
More pointedly,
phase $(Q)$ is associated to the fixed point $p^* \approx 1$ and phase
$(U)$ to $p^* \approx 1/2$. In Fig.\ \ref{fig:quasi1} we illustrate the 
dependence of the steady-state molecule frequencies $Y (i)$, given
in Eq.\ (\ref{bin}), on the  error rate $u$. Although
these results show a remarkable similarity to those obtained
with the original kinetics formulation of the quasispecies model
\cite{review}, the agreement is qualitative only: a
full analysis of the location of the error-threshold transition for the 
original model indicates that the predictions  based on the
recursion equation (\ref{rec_quasi}) are very inaccurate
\cite{Alves98}.

\begin{figure}
\vspace{9truecm}
\includegraphics{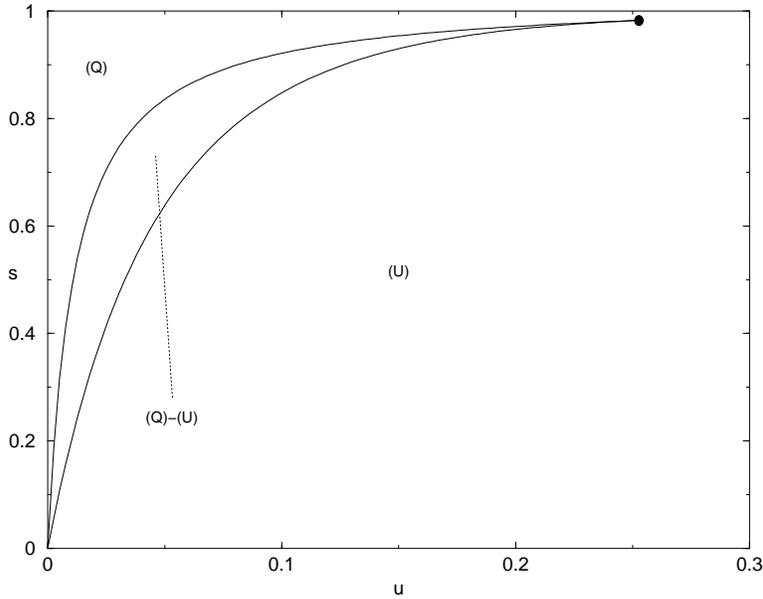}
\caption{Phase diagram for $N=10$  showing the regions of stability
of the quasispecies $(Q)$ and  uniform $(U)$ regimes. The discontinuous transitions
end at the critical point $u^c =0.251$ and $s^c= 0.983$.}
\label{fig:quasi2}
\end{figure}

In the present study the error-threshold transition corresponds
to the discontinuous transition between the phases $(Q)\!-\!(U)$ and $(U)$
(see Fig.\ \ref{fig:quasi2}). As in the previous models, the discontinuous
transition lines intersect and end at a critical point $(u^c,s^c)$ given
by \cite{Alves96}
\begin{equation}
u^c = 1 - \frac{1}{2} \left ( \frac{N+1}{N-1} \right )^2
\end{equation}
and
\begin{equation}
\frac{1}{1-s^c} = 1 + 2^N \left ( \frac{N-1}{N+1} \right )^N 
~\frac{ \left ( N-1 \right )^2 - 4 N}{N^2 -1} .
\end{equation}
It must be noted, however, that the rich interplay between phases
$(Q)$ and $(U)$ depicted in the phase diagram of Fig.\ \ref{fig:quasi2}
is a consequence of Wilson's trait group framework, in which the molecules
are disassembled and then randomly assembled during the life cycle  of
the population. Clearly, the use of that framework  is   
inadequate in the context of the quasispecies model, in which the
molecules and not the monomers are the self-replicating entities.

%
\section{Conclusion}\label{sec:level6}
%

In  this paper we have basically attempted to re-interpret
and unify several models dealing with the evolution of altruistic
behavior 
\cite{Michod,Roberta,DPS} in a single framework, namely, 
the `extended' Wilson's structured deme model of group selection \cite{Wilson}.
In doing so, we have carried out a thorough analysis of the steady-state 
regime of a model for the evolution of enzyme production proposed originally
by Michod \cite{Michod},  without resorting 
to the mean-field approximation implicit in Wilson's concept of average 
subjective frequencies \cite{Wilson}.
Furthermore, the effect of synergism  (i.e., division of labor)  was considered 
by assuming that the presence of altruists accrues benefits only to groups 
containing some minimal number of that type of  individual, following
thus Donato's alternative group selection model 
\cite{Roberta,DPS}.
In particular, we have obtained the 
phase-diagrams showing the regions of stability of the altruistic and 
non-altruistic regimes. A particularly relevant result is
the finding of a regime of stable coexistence within a same group
of altruists and non-altruists
which, though expected from observation, is not predicted by the
Island group selection models \cite{Eshel,Aoki,Ana}. 
We have also  identified
a recently proposed variant of
the quasispecies model, in which the macromolecules are viewed as vehicles
for the self-replicating monomers \cite{Alves96},  as a particular realization
of the `extended' Wilson's structured deme formalism presented in this paper.
In addition, we have found
that taking into account the possibility
of mutations leads to interesting qualitative changes on the 
steady-state regime of the model dynamics as, for instance,  
the appearance of critical points in the phase-diagrams of the models.
In particular, we have shown that there is
a value of the mutation rate $u_e$ (see Figs.\ \ref{fig:Mich5}
and \ref{fig:syner2}) above which the selective pressures are no longer
operative, in the sense that it is no longer possible to distinguish
between the altruistic and  the non-altruistic regimes. 

To conclude, we note that in the prebiotic context error-prone replication 
(mutation) has played a crucial role in revealing the 
limitations of non-cooperative molecular systems, such as Eigen's quasispecies
model, to function as efficient 
information integrators \cite{Eigen,review}. Furthermore, it was shown recently 
that  mutation can have disastrous 
effects over the stability of altruistic demes in the more traditional 
Island formulation of group selection theory \cite{Ana}. In view of this, 
mutation should not be viewed as merely another complication to be added to 
a model, but as a basic test for probing the robustness of any model 
of  integration  of information in prebiology, this being thus the  main
motivation for the present contribution.

\bigskip

\bigskip

The work of J.F.F. is supported in part by Conselho
Nacional de Desenvolvimento Cient\'{\i}fico e Tecnol\'ogico (CNPq)
and  Funda\c{c}\~ao de Amparo \`a Pesquisa do Estado de S\~ao Paulo 
(FAPESP), Proj.\ No.\ 99/09644-9. D.A., P.R.A.C. and A.T.S. 
are supported by FAPESP.


\end{document}